\documentstyle[12pt]{article}
\begin{document}
\begin{center}
 \Large {\bf  Thermal properties of a rotating nucleus in
a fluctuating mean field approach}\\
\vspace{0.75 true cm}
\large { \bf B. K. Agrawal$^1$ and A. Ansari}\\
\small {\bf  Institute of Physics, Bhubaneswar 751005, India}\\
\vspace{0.75true cm}
\end{center}
\footnotetext[1]{e-mail: bijay@iopb.ernet.in}


\vskip 1.0 true in
\noindent {\small {\bf Abstract:} The static path approximation to the
 path integral representation of  partition function provides a natural
microscopic basis to
deal with thermal fluctuations around mean field configurations.
 Using
this approach for one-dimensional cranking Hamiltonian with quadrupole-
quadrupole interaction term we have studied a few properties like
energy, level density, level density parameter($a$) and moment of
inertia as a function of temperature and spin for $^{64}Zn$ taking
it as an illustrative example.
We have also investigated the effects of variation  in interaction strength
on the level density and the parameter $a$
as a function of temperature.
The moment of inertia, $\cal I$ versus
rotational frequency, $\omega$ plot shows a sudden rise in the value of $\cal
I$
due to rotation alignment of $0g_{9/2}$ orbitals at $\omega\approx 1.0$ MeV
for a small temperature T $\sim 0.5$ MeV. At high T $\sim$ 2.0 MeV about
40-45$\%$ of each  angular momentum  is generated by
alignment  of $0g_{9/2}$ orbitals with an
interesting result that at $\omega\sim 1.0$ MeV and spin J $\sim$
16 the moment of inertia has almost a constant, temperature independent
value.

\noindent  PACS number: 21.10.Ma}

\newpage
\begin{center}
\section {\bf Introduction}
\end{center}
\label{Introduction}
Heavy ion fusion reactions deposit energy
into a nucleus   \cite{{Crema},{Bertsch}}
in such a way that it is shared by various intrinsic and collective
degrees of freedom. The
excited nuclei thus produced may keep their
energy long enough to reach internal statistical equilibrium[3].
The equilibrated system takes a time of the order of $10^{-21}
-10^{-19}$ sec.  \cite{pa} to decay by particle emissions, depending on the
excitation energy. During this time the intrinsic excitations are
present as thermal excitations and they may be observed by the
emission of photons. A quantitative interpretation of the $\gamma-$decay
rates, and the nuclear structure information  contained in,
consequently depends
on the nuclear level density. In addition to level density one
also studies a few other interesting features at finite temperature
and spin, e.g., shape transitions, collapse of proton and
neutron pairing correlations, moment of inertia and  rotational
damping[4-9] etc.

\par   The problem of level density has been a subject of interest,
theoretically as well as experimentally.
 Theoretically, it provides an important
basis to test the validity of the approximations
to many-body problem. The level density has
been extensively studied semi-classically using Thomas-Fermi approach
including quantal effects involving quite a few temperature dependent
parameters   \cite {{Shlomo prc44},{Shlomo plb252}}. It has also
been studied   microscopically
using finite temperature mean field theories, like, Hartree-Fock or
Hartree-Fock Bogoliubov
  \cite{Goodman npa352}.
The success of these mean field theories is based ,to some extent, on the
symmetry breaking. The symmetry breaking allows a considerable
enhancement in variational Hilbert space and thereby includes
various correlations appropriately.
Ultimately these  broken symmetries can be restored
by using standard symmetry projection techniques   \cite{Ring}.
But each symmetry breaking introduces thermal and quantal
fluctuations in the related
degree of freedom, e.g. the nuclear orientation fluctuations caused
by the breaking of rotational symmetry.
Since, a nucleus is a finite system,  these fluctuations play a
vital role in understanding its dynamics.
It has been
shown explicitly by Egido  \cite{Egido} that the quantal
fluctuation at lower temperature
dominates over the thermal fluctuation, but as temperature increases,
thermal fluctuation grows faster  whereas quantal fluctuation dies
out. Recently, Alhassid and Bush\cite{Alhassid prl65} have included
the effect of orientation fluctuation in Landau theory of phase
transition and find that it is required to explain the observed
angular distribution for giant dipole resonance. On the otherhand, Goodman
\cite{Goodman prc39} explains the observed collectivity in
a hot rotating $^{168}Yb$
only when the thermal shape fluctuations are
included in the mean field.
However, the path integral
representation of partition function   \cite{Feynman}
provides a natural framework to deal with the thermal as well
as quantal fluctuations.  The exact path integral representation
of the partition function can be obtained by exploiting the
Hubbard-Stratonovich transformation\cite {{Hubbard prl3},{Stratonovich
Doklady}}.
This transformation introduces
the path integration over some auxilliary field variables which
are coupled to the one-body or the pairing density matrices and
permits the linearization of two-body interaction with respect to
these density matrices.
One makes various approximations
on these field variables to obtain mean field solution or a solution including
quantal and thermal fluctuations around mean field. In static
path approximation
(SPA)\cite {{LAB prl61},{LB prc39},{Alhassid prc30}}
these fields are restricted to static paths (or static
single-particle potential ) which describes the motion of A-nucleons
in a fluctuating mean field.
As expected, it is found that the results obtained within SPA  at high
temperature are quite
close to one obtained by exactly solvable models.
Furthermore, when small amplitude quantal correction or RPA correlations
  \cite {{Broglia prc42},{Broglia ap206}} are also included there is a
remarkable improvement at low temperatures.

\par The static path approximation for nuclear partition function
with quadrupole-quadrupole interaction Hamiltonian allows  a nucleus
to span entire collective space characterized by the deformation parameters
$\beta$ and $\gamma$. So, to get some meaningful information, it is required
to restore the
rotational symmetry by using three-dimensional angular momentum
projection at each point in the $\beta-\gamma$ plane\cite{Ansari prl}. This way
of restoring symmetry needs the evaluation of five dimensional integrations,
i.e, integration over three Euler angles and  two deformation
parameters, $\beta$ and $\gamma$. However, if one is not looking at quantities
very sensitive to orientation
fluctuations, then for a qualitative study cranking may be a reasonable
proposition.  Already in the previous work\cite{BA prc} we have
seen that variation of moment of inertia as a function of rotational
frequency at finite temperatures comes out quite encouraging.
. In the present work we have
used this formalism to constrain the average spin and studied
the spin and temperature dependence of various quantities
for $^{64}Zn$, including nuclear level density.
Here we must mention that, as is well known, $\omega=0$ does not really
corresponds to $J=0$ and only at high spins like $J\ge8$
the angular momentum constraint implies a most probable spin value.
 Though this method does not affect
the dimensionality of integrations,
the dimension of Jacobian appearing in the expression for level density
becomes four instead of three.

This paper is organized as follows:
In the next section we present briefly the theoretical framework
including basic expressions required for numerical computations,
following mainly ref.  \cite{{LAB prl61},{LB prc39}}. Section 3
contains some numerical details and discussions
of the results. Finally our main conclusions are presented in Section 4
 along with a brief summary and prospects for subsequent
calculations.

\begin{center}
\section {\bf Theoretical framework}
\end{center}
\label {theory}
\par As already mentioned above, we follow the path integral representation
of partition function within  static path approximation (SPA)
for quadrupole interaction as described by Lauritzen {\it et al} \cite
{LAB prl61}. Only
briefly we outline
the formulation listing essential equations. The
grand canonical partition function for a system rotating about
intrinsic x-axis is
\begin{equation}
{\cal Z}=Tr\>e^{-(\hat H^\omega-\mu_p\hat Z-\mu_n\hat N)/T}
\end{equation}

\noindent where,  $\omega$ is the rotational frequency and
\begin{equation}
\hat H^\omega=\hat H-\omega \hat J_x
\end{equation}
\noindent  $\hat J_x$ being the x-component of the angular momentum
operator $\hat J$. Considering a quadrupole interaction Hamiltonian,
written as
\begin{equation}
\hat H=\hat H_o-{1\over 2}\>\>\chi\> \hat Q\> .\> \hat Q
\end{equation}
\noindent where $H_o $ is the unperturbed spherical part and
\begin{equation}
\hat Q_\mu=r^2\>\>Y_{2\mu}
\end{equation}

\noindent is the quadrupole moment operator. The path integral representation
of partition function in the
SPA is given by

\begin{equation}
{\cal Z}(\mu_p,\mu_n,\omega,T)=Tr\hat {\cal D}
\label {partition function}
\end{equation}
\noindent where,
\begin{equation}
\hat{\cal D}=4\pi^2({\alpha\over 2\pi T})^{5/2}\int_o^\infty
\beta^4d\beta
\int_o^{\pi/3}\mid sin3\gamma\mid d\gamma  e^{-{\alpha\beta^2\over
2T}}e^{-(\hat H^{\prime
^\omega}-\mu_p\hat  Z-\mu_n\hat  N)/T}
\label {static path operator}
\end{equation}
\noindent is the static path statistical operator for quadrupole-quadrupole
interaction Hamiltonian. The one-body operator  $\hat H^{\prime\omega}=
\sum_ih^{\prime\omega}(i)$
is a  Nilsson type deformed  mean field Hamiltonian
\begin{equation}
\hat h^{\prime\omega}= h_o-\hbar \omega_o\>\beta\> {r^2\over b^2}
[cos\gamma\> Y_{2\>0}+
{1\over \sqrt 2}\>sin\gamma\> (Y_{2\>2}+Y_{2\>-2})]-\omega\hat J_x
\label {deformed hamiltonian}
\end{equation}

\noindent with $h_o$ representing here the spherical basis space
single-particle energies defined with respect to an appropriate inert core.
The value of $\hbar\omega_o=41/A^{1/3}$ MeV and $\alpha=(\hbar\omega_o)^2/
{\chi b^4}$ with $\chi b^4$ =70 $A^{-1.4}$ MeV as given by Baranger
and Kumar  \cite{BK npa110}.
\par The chemical potentials for proton and neutron are determined from the
relations

\begin{equation}
Z,N=T{\partial\> \over \partial\>  \mu_{p,n}}\> ln{\cal Z}(\mu_p,\mu_n,\omega,
T)
\label {number equation}
\end{equation}
\noindent Similarly, the desired value of average angular momentum $\sqrt
{J(J+1)}$ is obtained by adjusting $\omega$ such that
\begin{equation}
\sqrt{J(J+1)}\> =\> <J_x>\> =\> T{\partial \over \partial \omega}
ln{\cal Z}(\mu_p,\mu_n,\omega,T)
\label {spin equation}
\end{equation}
\noindent The energy as a function of temperature at a fixed number of
particles and
spin is given by
\begin{equation}
E(T)=T^2{d \over d T}\> ln{\cal Z}\>+\>\mu_p \>Z+\>\mu_n \>N+\>\omega <J_x>\>
\label {energy}
\end{equation}
\noindent and moment of inertia ${\cal I}$ is defined as
\begin{equation}
{\cal I}={<J_x>\over \omega_{\scriptscriptstyle{J}}}
\label {moment of inertia}
\end{equation}
\noindent where $\omega_J$ is such that the constraint Eq.
(\ref {spin equation})
is satisfied.
\noindent The nuclear level density is evaluated from the inverse Laplace
transform of the partition function  ${\cal Z}$. For the fixed number
of protons and neutrons, in the saddle point
approximation,  it is given by \cite{BM 75}
\begin{equation}
\rho(E,J)={e^S\over (2\pi)^2\>\>D}
\label {level density}
\end{equation}
\noindent  where,
\begin{equation}
S =(E-F)/T
\label {entropy}
\end{equation}
\noindent is the entropy  and the free-energy F is given by
\begin{equation}
F=-Tln{\cal Z}+\mu_p Z+\mu_n N+\omega <J_x>\
\label{free energy}
\end{equation}
\noindent The quantity $D$ is the square root of Jacobian,
i.e $\sqrt {\it J}$ with
\begin{equation}
\it {J}={\partial (E,Z,N,J)\over \partial (\beta,\alpha_p
,\alpha_n,\lambda)}
\label{Jacobian}
\end{equation}

\noindent where, $\beta=1/T$, $\alpha_{p,n}=-\mu_{p,n}/T$ and
$\lambda=-\omega/T$.

\par Finally, it may also be useful to
calculate an effective level density parameter  which is often used
to connect the intrinsic excitation energy with a temperature
\begin{equation}
E^*(T)=a_{eff} T^2
\end{equation}
\noindent That is $a_{eff}=(E(T)-E(T=0))/T^2$. It is now obvious that
the numerical value of $a_{eff}$ will strongly depend on the correct
evaluation of the binding energy at T=0. On the otherhand SPA is not
applicable in the T=0 limit. One way this ambiguity
can easily be  removed is by taking a derivative of $E^*(T)$ with
 respect to T, so that now (ignoring the dependence of $a_{eff}$
on T locally).
\begin{equation}
a_{eff}={1\over 2T}{dE\over dT} \> \> \>\quad\quad\quad\quad
\quad\quad\quad\quad (\>T>0\>)
\label {a effective}
\end{equation}
\par As it will be seen in the next section, we have also used another
expression for $a_{eff}$ in terms of entropy to compute its values
\begin{equation}
a_{eff}={S\over 2T}.
\end{equation}

\begin{center}
\section  {\bf Numerical details and results}
\end{center}
\label{results}

In this section we give
 some details of the numerical calculations performed,
and present the main results for a   nucleus $^{64}Zn$. We have studied
the spin and temperature dependence of various quantities like,
energy, level density parameter, level density and moment of inertia.
In order to have  a reasonable number of active valence particles we have
chosen Z=20 and N=20 (i.e  $^{40}Ca$) as an inert core. Thus we have 10
protons and 14 neutrons each in  30 orbitals spanning the model basis space up
to
$0g_{9/2}$.
 This model space would be very reasonable for zero temperature calculations
 even for high spin states. However, at  high temperatures this would lead to
truncation
 effects. We have tested that our results are reliable up to T=2 - 2.5 MeV.
 The spherical basis single-particle(sp) energies are
-14.4, -10.2, -8.8, -8.3 and -4.4 (all in MeV) for the orbitals
$0f_{7/2}$, $1p_{3/2}$, $0f_{5/2}$, $1p_{1/2}$ and
$0g_{9/2}$, respectively.
 These values are precisely those given by Lauritzen and Bertsch
\cite {LB prc39}.
The matrix elements
of the sp Hamiltonian (\ref {deformed hamiltonian})  can be easily
calculated
following Baranger and Kumar\cite {BK npa110}. However, we should
point out that the matrix elements of
$r^2$ for the basis states beyond one major shell need to be reduced
(renormalized) as is done in \cite {BK npa110}.  More realistic would
be to use a radial function f(r) like that employed in
ref.\cite {LB prc39},
particularly when the basis space spans beyond one or two major shells.

 \par Obviously as a first step of the calculations the deformed sp
Hamiltonian (\ref{deformed hamiltonian})
is diagonalised in the basis space
at grid points in the $\beta-\gamma$ plane for a fixed value of $\omega$
so that a
numerical integration( 12 point Gaussian in the $\beta$ space with
$\beta_{max}=0.5$ and  8 point Gaussian for $\gamma=0-60^o$) in
Eq. (\ref {static path operator})
can be performed to evaluate the partition function $\cal Z$. One may
compute the r.h.s. of Eq. (\ref {static path operator}) on a number of mesh
points $\mu_p, \mu_n$,
T and $\omega$  so that later on a required value of $\cal Z$  or its
derivative at
any $\mu_p, \mu_n$, T and $\omega$ point could be computed numerically
using multidimensional interpolation.
However, we take the required differentiation $\partial  /
\partial  \mu$
, $\partial /\partial T$ or $\partial /\partial \omega$  directly inside the
integration sign on the
r.h.s. of Eq. (\ref{static path operator})
 and then the ($\beta,\gamma$) integrations are performed
for each quantity separately.
\subsection {ENERGY}
\label {energy spectrum}

\par In Fig. \ref {E vs T} we have displayed the variation of energy (Eq.
(\ref {energy}) as
a function of temperature for a few spin values  J=0,4,8 and 16. Experimentally
\cite{B. Singh} for the yrast (T=0) J=2 and 4 states the excitation energies
are $E_2=0.992$ MeV and $E_4=2.307$ MeV, respectively. But in the
present calculation these come out quite compressed (our lowest
T=0.3 MeV). Of course, at T$\le 0.5$ MeV the absence of pairing should be
one of the main reasons  for this compression. However, for T$\ge 1.0$ MeV
our results  should be realistic. We also notice that $E_I-E_I(T=0)$ is the
largest for J=0 and this difference would have been even still larger
particularly for the low spin states had the pairing correlations
been included. The temperature dependent CHFB mean field calculations
of Egido et al\cite {Egido npa451} for $^{164}Er$ also show a similar
behaviour.

\par At the first sight the plot of Fig. \ref{E vs T} in the $T\rightarrow 0$
limit appears like a rotational one. But actually it is in between rotational
and vibrational, e.g. $E_6-E_4=0.921$ MeV and $E_8-E_6=1.42$
MeV. Besides the absence of pairing, the compression of the spectrum,
particularly at high spins, is caused by the rotation alignment effects.
At J=2 and T=0.5 MeV the occupation of $0g_{9/2}$ orbitals is zero.
But at higher spins  and/or higher temperatures $0g_{9/2}$ orbitals
get partially occupied and help in easy generation of angular momenta
through alignment at low energy expenses. More on this will be discussed
latter on.

\par Though in SPA there is integration over the deformation parameter
($\beta,\gamma$) space, looking at the surface plots of the free energy like
quantity defined as
\begin{equation}
{\it f}(\beta,\gamma)={\alpha\beta^2\over 2}-ln[tr\> exp(-(\hat H^{\prime
\omega}-\mu_p Z-\mu_n N)/T)]+\mu_p Z+\mu_n N+\omega <J_x>
\end{equation}

\noindent may give some insight on the
average shape evolution as a function of temperature and spin.
We have choosen just three contour plots in the $\beta-\gamma$ space:
Figs. 2a, b and c\ for T=0.5 MeV, J=0; T=0.5 MeV, J=8; T=2.0 MeV, J=16,
respectively. The energy difference between the successive contour
lines is 1.0 MeV and the numbers (energy in MeV) on a few of the lines give
the idea of the free energy surface around the minimum  value.
The solid dots indicate the minimum point with equilibrium value
of deformation parameters ($\beta_o, \gamma_o$). Large spacings near
the minimum free energy lines indicate the extent of shallowness
in energy and importance of shape fluctuations. Comparision of Figs. 2a and b
shows that even at low temperature T=0.5 MeV and at J=8 the most probable
(mean field or Hartree) shape has changed to oblate from prolate at
J=0. At high spin J=16 and T=2.0 MeV also the shape is oblate (Fig. 2c),
whereas at J=0 and T=2.0 MeV the mean field shape is spherical[20]. The
value of $\beta_o$ in all the three cases is about 0.2. In Fig. 1  there is
another curve (dashed) for J=16 evaluated at the most probable value of
deformation parameter( see Fig. 2c) $\beta_o$ and $\gamma_o$. Though the
difference between
the two J=16 curve is not large ($\delta E \sim 500$ keV), it is not
negligible particularly at higher temperatures. Even such small differences
may be indicative of large fluctuation effects in other dynamical
properties such as transition densities.

\subsection { LEVEL DENSITY AND LEVEL DENSITY \break PARAMETER '$a$'}
\label {rho}
\par At high temperatures the $\gamma$ decays  are predominantly
statistical in nature. Therefore, it is impractical to resolve these
$\gamma$-rays
to study the individual band structures. For a statistical analysis of these
decay properties one needs excitation energy (temperature) and spin dependent
level densities. Approximate spin dependent level density can be computed
using the expression (\ref {level density}). The variation of ln$\rho$ versus T
is shown
in Fig. \ref {ln(rho)} for J=0,4,8,16 and 28. The J=0 curve is of course the
$\omega=0$
no cranking result as in  ref.\cite{BA prc}. Looking at J=4,8 and 16
curves  there are two trends visible. One is that ln$\rho$ is increasing with
the increase of T for all the J values considered and the other is that
at low temperature(T=1.0 MeV) the value of ln$\rho$ is lower for the
smaller J value and by T=2.0 MeV this trend gets reversed. Also at T=2.5
MeV the J=16 curve shows a tendency of saturation. The above mentioned
reversal may be indicative of the limited basis space at high T and high spin.
Just to verify this argument we have repeated the calculation of
ln$\rho$ for J=28, as shown in Fig. \ref{ln(rho)} by a solid line
with cross($\times$) marks.
Slight flattening tendency of this curve for T$>$ 2.0 MeV once again
may be  indicative of
the limited basis space.

\par It should be useful  to study the spin and temperature dependence of the
level density parameter $a$ which is often used to compute the level density
employing
semi-empirical or phenomenological expressions. Curves in Fig. \ref{E vs T}
show
a rather smooth variation of energy as a function of T. However, their
slopes are changing,  being very small for T$<$ 1.0 MeV and
almost a constant for T$>$ 1.0 MeV. From Eq. (\ref {a effective}) it is
imployed that
$a$ may be very small at low value of T. In Fig. 4 we have shown
the inverse level density parameter K=A/$a_{eff}$ as a function of
temperature for some selected spins J=0, 4, 8 and 16. As in the
absence of Fock term the slope of E$^*$ vs T curve is very small for T$\le$
0.5 MeV the corresponding K-curves in Fig. \ref {K vs T} show negative slopes.
For T$\ge$ 1.0 MeV  K increases with the increase of T for all spins.
In our previous work\cite {BA prc}  without the consideration of spin
($\omega=0$)
change in K was slower; for T=0.5 MeV to 1.5 MeV,  increase in K is
roughly by 8$\%$. Here we must mention that presently the basis space
is larger for neutrons as the inert core size is reduced to $^{40}Ca$
whereas earlier\cite{BA prc}  it was $^{48}Ca$. So, for example, at T= 1.0 MeV
now K is smaller by about 10$\%$ and this is understood to be caused
by the increase in collectivity, there being more number of active
particles. In view of the present cranking calculations J=8 and 16 cases
should be taken more seriously and for these there is no  flat (constant
value of K) region. For T$>$ 2 MeV  the increase is even faster which may
be an indication of the basis truncation effect.

\par  As mentioned in Sec. 2 the quadrupole interaction strength
$\chi$ (in MeV) = c/A$^{1.4}$ with c=70 has been used in our computations.
Without much justification we have taken it to be just the same as used
by Baranger and Kumar\cite{BK npa110} for the rare-earth nuclei. Then we have
considered
 two more values of c=65 and 75 in order to investigate the
effect of interaction on level density(only $\omega=0$ case) . Of course, we
know that
higher the value of c interaction is more attractive and it should lead to
lowering in energy and a more deformed system. In confirmity with this
we do get almost parallel running curves for c=65, 70 and 75 in the E-T
plane. But its  effect on the value of ln$\rho$ (Eq. (\ref{level density}) and
K
(Eq. (\ref {a effective}))
does not seem to be a priori obvious. We have computed these quantities
and enumerated them in Table 1 as a function of temperature with
c=65,  70 and 75. First we consider the value of ln$\rho$.
At T=1.9 MeV we notice that ln$\rho$ is practically independent of $\chi$.
At T$<$ 1.9 MeV ln$\rho$ seems to be increasing with the increase
in the value of $\chi$, so much so that at T=1.1 MeV $\rho$ has
increased by  about 25$\%$  in going from c=65 to 75 (an increase of about
15$\%$). Though actual numbers do not precisely support, we feel that
at T$\ge$ 2 MeV ln$\rho$ is not too sensitive to the interaction strength.
At T= 2.5 MeV we notice decrease in the value of ln$\rho$ with
increase of $\chi$ which may be a manifestation of limited basis space. The
value
of K seems to be slightly increasing with the increase of $\chi$, that is the
value of the level density parameter, $a$ is decreasing with the increase
of the value of $\chi$. It does not look convincing but nontheless
may be true.

\par On the other hand the values of $K_S$ in Table 1 indicate that using
Eq. (18) $a_s
= S/2T$  is slightly  higher for the large value of  $\chi$ at low temperatures
and by T=2 MeV it becomes insensitive to about 15$\%$ increase in the value of
$\chi$. An important point to note here is that the
magnitude of $K_S$ is much smaller compared to the value of K
and it is around the prevalent value in use in the semi-empirical
calculations in the literature, that is
about 8 to 10. In this sense the definition of the level density parameter
as S/2T seems more reasonable as it is also consistent with the variation
of ln$\rho$ presented in Table 1. Before proceeding to next section we
may still note that the rate of increase in the value of $K_S$ with
T is rather large (not at all a constant) and is similar to that of K.

Finally, we conclude this section with an important remark in support
of our above investigation for the variation of ln$\rho$ with $\chi$.
In a very recent publication Alhassid and Bush \cite {Alhassid npa549} have
studied  the  nuclear level density in SPA applied
to an exactly solvable SU(2) model. Within this model study
they do find that ln$\rho$ increases continuously with the increase in
interaction strength at low excitation energy and at high excitation
energy its value appears almost insensitive to the interaction strength
(Fig. 7 in ref. \cite {Alhassid npa549})
\subsection { MOMENT  OF INERTIA}
\label {inertia}

According to Eq. (11) we define moment of inertia, $\cal I$ as a ratio
of the angular momentum $\sqrt {J(J+1)}$ to the cranking frequency $\omega_{
\scriptstyle J}$.
This way the moment of inertia is not a parameter solely describing
the collective rotation of a nucleus. However, this makes it a more
interesting parameter as its variation with $\omega$ at various temperatures
can
depict the effect of interplay between collective and sp degrees of freedom,
i.e  rotation alignment.
Therefore, like a backbending plot\cite {Stephens rmp47} we have shown in Fig.
5 the
variation of $\cal I$ as a function of $\omega$ at four values of
T between 0.5 and 2.0 MeV. On each curve the points indicate the value of
J=2, 4, 6 ..  such that the constraint relation (9) is exactly satisfied. The
horizontal dashed curve represents  the value of rigid body moment of inertia
at T=0 and the deformation $\beta=0.2$ and  $\gamma=0$. At a small temperature,
T=0.5 MeV plot shows a sudden rise of $\cal I$ for $\omega >$ 0.5 MeV
implying the generation of spin by alignment of a high-j sp orbital
along the rotation axis. As the Table 2 shows, the orbital
$0g_{9/2}$ is unoccupied at J=2 and T=0.5 MeV. But with a slight
increase of $\omega$ the level $0g_{9/2}$ start getting partially occupied
and contribute to the generation of total angular momentum. So much so that
at J=16 the $0g_{9/2}$ orbitals, still with less than two particle in it,
contribute about 40$\%$. Denoting the contribution of particles in
$0g_{9/2}$ orbitals as aligned angular momentum, $j_a$ and the
rest as collective,  $J_{coll}$ one can write

\begin{equation}
J=J_{coll}+j_a
\end{equation}

\noindent In Table 2 we have listed the value of $j_a$, percentage
contribution of $j_a$ to total angular momentum $<J_x>$ and
the number of particles in $0g_{9/2}$ at two temperatures T= 0.5
and 2.0 MeV. Only  a few values of angular momenta are given, out of which
J=16 is of a special interest. At J=16 the value of the moment of
inertia ${\cal I} = <J_x>/\omega_{\scriptstyle J}$ remains almost independent
of temperature(see Fig. 5).
Furthermore at T=2.0 MeV the occupation
of $0g_{9/2}$ orbitals  has substantially increased compared to the
T=0.5 case and keeps growing with the increase of the value of J.
However, the percentage contribution of $j_a$ at all the spins
is almost the same. For $J<16$  it is a
temperature induced alignment whereas for $J\ge 16$ it is the usual
rotational alignment\cite{Stephens rmp47}
 present even at zero temperature.
We may point out that with two protons and two neutrons in ${0g_{9/2}}$
it can contribute maximum $j^{max}_{a}=16$. As seen from Table 2 both
protons and neutrons are contributing to the rotation aligned component
$j_a$ which may be termed as a coherent rotation alignment and this is
stronger at a higher temperature.

\begin{center}
\section {\bf  Summary and conclusions}
\end{center}
\label{summary}

Employing a one dimensional cranking Hamiltonian with
quadrupole-quadrupole interaction term in the static path approximation
we have studied the spin and temperature dependence of energy, level density,
level density parameter and moment of inertia of $^{64}Zn$. The angular
momentum is, of course,  conserved  only on the average according to the
constraint given
in Eq. (9).The level density parameter $a$ is calculated using two
expressions: dE/dT = 2$a$T and S=2$a$T. The numbers obtained using latter
relation seem to be more in agreement with the empirical
values. Effect of variation of the interaction
strength on the thermal properties is also investigated and it seems that
at high temperature T$\ge $ 2.0 MeV the dependence is rather
weak. Within this investigation the variation of $a$ or K=A/$a$
as a function of T again appears to support the S=2$a$T definition of $a$
so that its behaviour is consistent with that of the level density
$\rho$.

\par The variation of moment of inertia as a function of spin
as well as temperature is studied with a definition ${\cal
I}=<J_x>/\omega_{\scriptstyle J}$.
At a given low temperature T$<$ 2.0 MeV $\cal I$ increases with the
increase of J and becomes almost a constant at T=2.0 MeV. At
T$<$ 1.0 MeV and high spin J$\ge$ 16 about 40$\%$ of the total spin value  is
generated by the
alignment mechanism of $0g_{9/2}$ orbitals. On the other hand at T=2.0 MeV
the alignment of $0g_{9/2}$ orbitals contribute about 40$\%$ to all the angular
momenta.
We are planning to include J-dependence according to Eq. (8)
of ref. [19] so that the validity of the present
cranking results can be ascertained.
\pagebreak

\begin{figure}[pt]
{\bf Figure Captions}

\caption { Energy versus temperature for $^{64}Zn$ at various spins,
J=0, 4, 8 and 16. The dashed curve represents the energy for J=16 at
different temperatures evaluated with most probable or equilibrium
value of deformation parameters $\beta$ and $\gamma$.\label {E vs T}}

\caption {Contour maps for free energy $\it f$($\beta,\gamma$) in
$\beta-\gamma$
plane for $^{64}Zn$ at different spins and temperatures, (a) J=0 and T=0.5 MeV,
(b) J=8 and T=0.5 MeV and (c) J=16 and T=2.0 MeV. Each contour represent
a path in $\beta-\gamma$ plane for constant $\it f$($\beta,\gamma$) which
differs by 1.0 MeV in magnitude for the adjacent contours. The solid
dot indicates the point at which ${\it f}(\beta,\gamma)$
 is minimum.\label {contour}}
\caption { Logrithmic variation of nuclear level density $\rho$ for
$^{64}Zn$ as a function of temperature at spins J = 0, 4, 8,16 and 28.
.\label {ln(rho)}}
\caption {Spin and temperature dependence of inverse level density parameter
$K=A/a_{eff}$ for $^{64}Zn$. The negative slope at lower spin and temperature
indicates the lack of contribution due to pairing correlations and
Fock energy. \label {K vs T}}

\caption { Systematics of moment of inertia ${\cal I}$ vs rotational
frequency $
\omega$ for $^{64}Zn$ at finite temperatures, T = 0.5 - 2.0 MeV.
Points on these curves satisfy the angular momentum constraint
({\it see} Eq. (9)).
The sudden rise in $\cal I$ at $\omega\approx 1.0$ MeV for T=0.5 MeV
shows a rotation alignment of $0g_{\scriptstyle 9/2}$ orbitals. The horizontal
dashed line represent the rigid body moment of inertia ${\cal I}_{rig}$
for $\beta=0.2$ and $\gamma=0$.
\label {I vs w}}
\end{figure}
\large
\begin{table}[pt]
\caption{ The values of ln$\rho$, K and K$_S$ with no cranking
at different temperature
for interaction strength with $c_1$, $c_2$, $c_3$ corresponding to c=65, 70
and 75, respectively in $\chi=cA^{-1.4}$.}
\vspace {0.5in}
\begin{center}
\begin{tabular}{|c|c|c|c||c|c||c|c|}
\hline
\multicolumn{1}{|c|}{T }&
\multicolumn{3}{|c||}{ln$\rho$}&
\multicolumn{2}{|c||}{K (MeV)}&
\multicolumn{2}{|c|}{K$_S$ (MeV)}\\
\cline{2-8}
\multicolumn{1}{|c|}{(MeV)}&
\multicolumn{1}{|c|}{$c_1$}&
\multicolumn{1}{|c|}{$c_2$}&
\multicolumn{1}{|c||}{$c_3$}&
\multicolumn{1}{|c|}{$c_1$}&
\multicolumn{1}{|c||}{$c_3$}&
\multicolumn{1}{|c|}{$c_1$}&
\multicolumn{1}{|c|}{$c_3$}\\
\hline
    0.5  & 7.32  & 7.43 &  7.59 & 13.51 & 12.88 &  6.21 &  6.15 \\
\hline
    0.9 &  9.76 &  9.89 & 10.13 & 12.12 & 12.34 &  8.17 &  8.00\\
\hline
    1.1  &11.43  &11.56 & 11.66 & 12.35 & 12.71 &  8.69 &  8.59 \\
\hline
    1.5  &14.73  &14.81 & 14.91 & 13.68 & 14.17 &  9.50 &  9.46 \\
\hline
    1.9  &17.73  &17.74 & 17.73 & 16.34 & 16.84 & 10.26 & 10.26 \\
\hline
    2.1  &19.03  &19.03 & 19.01 & 18.48 & 18.91 & 10.71 & 10.71 \\
\hline
    2.5  &21.25  &21.23 & 21.21 & 26.35 & 26.11 &  11.82 &  11.59 \\
\hline
\end{tabular}
\end{center}
\end{table}
\pagebreak
\begin{table}[pt]
\caption{
The value of rotation aligned angular momentum  $j_a$, percentage
contribution of $j_a$ to total angular momentum $<J_x>$ and
the number of particles in $0g_{9/2}$ at temperatures T= 0.5
and 2.0 MeV. }

\vspace {0.5in}
\begin{center}
\begin{tabular}{|c|c|c|c|c||c|c|c|c|}
\hline
\multicolumn{1}{|c|}{J($\hbar$)}&
\multicolumn{4}{|c||}{T=0.5 MeV}&
\multicolumn{4}{|c|}{T=2.0 MeV}\\
\cline{2-9}
\multicolumn{1}{|c|}{}&
\multicolumn{1}{|c|}{$j_a$}&
\multicolumn{1}{|c|}{${j_a\over J}\%$}&
\multicolumn{2}{|c||}{$0g_{9/2}$ occupation($N_{g_{\scriptstyle 9/2}})$}&
\multicolumn{1}{|c|}{$j_a$}&
\multicolumn{1}{|c|}{${j_a\over J}\%$}&
\multicolumn{2}{|c|}{$0g_{9/2}$ occupation($N_{g_{\scriptstyle 9/2}})$}\\
\cline{4-5}
\cline{8-9}
\multicolumn{1}{|c|}{}&
\multicolumn{1}{|c|}{}&
\multicolumn{1}{|c|}{}&
\multicolumn{1}{|c|}{Protons}&
\multicolumn{1}{|c||}{Neutrons}&
\multicolumn{1}{|c|}{}&
\multicolumn{1}{|c|}{}&
\multicolumn{1}{|c|}{Protons}&
\multicolumn{1}{|c|}{Neutrons}\\
\hline
2&0.11& 4.5&0.00&0.00 &1.05& 42.8 &0.42&1.07 \\
\hline
8&1.93& 22.7&0.00&0.38&3.56&42.0 &0.52&1.24 \\
\hline
16&6.19& 37.5& 0.32&1.02& 7.41&44.9&0.81& 1.63\\
\hline
20&7.74& 37.8&0.65&1.22&  9.01&43.9&1.00& 1.85\\
\hline
28&11.29& 39.6&1.07&1.81& 12.69 &44.5&1.45&2.35 \\
\hline
\end{tabular}
\end{center}
\end{table}

\pagebreak

\pagebreak

\end{document}